\documentclass[fp]{jpsj3}
\usepackage{cite}
\usepackage{txfonts}
\usepackage{amsmath}
\usepackage{amssymb}
\usepackage{graphicx}
\newcommand{\bi}[1]{
\ensuremath{\boldsymbol{#1}}}

\title{Energy Gap at First-Order Quantum Phase Transitions: An Anomalous Case}

\author{Junichi Tsuda, Yuuki Yamanaka, and Hidetoshi Nishimori}
\inst{Department of Physics, Tokyo Institute of Technology\\
Oh-okayama, Meguro-ku, Tokyo 152-8551, Japan} %\\

\abst{We show that the rate of closing of the energy gap between the ground state and the first
excited state, as a function of system size, behaves in many qualitatively different ways 
at first-order quantum phase transitions of the infinite-range quantum $XY$ model.
Examples include polynomial, exponential and even factorially-fast closing of
the energy gap, all of which coexist along a single axis of the phase diagram
representing the transverse field.
This variety emerges depending on whether or not the transverse
field assumes a rational number,
as well as on how the series of system size is chosen toward the thermodynamic limit.
We conclude that there is no generically applicable rule for relating the rate of gap
closing to the order of quantum phase transitions as is often implied in many studies,
particularly in relation to the computational complexity of quantum annealing in its
implementation as quantum adiabatic computation.}

\kword{quantum phase transition, quantum annealing, energy gap}

\begin{document}
\maketitle

\section{Introduction}
Quantum annealing is a generic algorithm to solve combinatorial optimization problems,
expressed in terms of the Ising model, using quantum fluctuations or quantum tunneling
\cite{KN,MN08,Finilla,DC,ST,SICbook}.
The system, typically the transverse-field Ising model with complex interactions,
is initially set to the ground state of a trivial quantum Hamiltonian, e.g., the transverse-field
term, and then is driven by the time-dependent Schr\"odinger equation toward the ground state of
the final Hamiltonian corresponding to the solution of a given combinatorial optimization problem.
A descendant of quantum annealing is quantum adiabatic computation \cite{FGGLL}, in which the 
system is assumed to follow the instantaneous ground state of a time-dependent quantum Hamiltonian.
We study this latter realization of quantum annealing in the present paper.

The initial and final states are quite different from each other, the former
being trivial and the latter highly non-trivial.
This implies that the initial and final states belong to different thermodynamic phases.
Therefore, after the thermodynamic (large-size) limit is taken, the system is expected
to undergo a quantum phase transition.

It is often the case that the energy gap between the ground state and the first excited
state closes at a quantum phase transition point, which causes difficulties because the
adiabatic condition of quantum mechanics states that the time scale to stay in the
instantaneous ground state is inversely proportional to the square of the energy gap.

One of the main interests in combinatorial optimization problems is the computational
complexity, i.e., the time necessary for a given algorithm to reach the solution as a function
of the system size (problem size).
The system size is usually large but finite, and thus our task amounts to the determination
of whether the gap closes very quickly, typically exponentially, as a function of
system size or relatively slowly or  polynomially.
It is generally believed, and is indeed the case in many instances,  that the gap
closes exponentially fast at first-order quantum phase transitions whereas it is
polynomial for second-order transitions.
Many researchers have therefore been attempting to determine the order of transitions
in quantum systems representing quantum annealing.
See Refs. \cite{SekiN,SeoaneN,BS,JKKMP} and references cited therein.

An interesting counterexample was presented by Cabrera and Jullien \cite{CJ} who showed that the
first-order quantum phase transition in the one-dimensional transverse-field Ising model accompanies
a polynomial closing of the energy gap if one imposes an antiperiodic boundary condition.
See also Ref. \cite{Laumann} for essentially the same result.
In the present paper, we give another quite unusual example where the energy gap
closes in widely different ways -- polynomial, exponential, and factorial -- depending strongly on
the value of the parameter in the Hamiltonian as well as on the choice of the sequence of
system size toward the thermodynamic limit.

In the next section, the model system is described and its thermodynamic behavior is analyzed.
\S 3 constitutes the main body of this paper, where the behavior of the energy gap
for finite-size systems is studied in detail.
The final section is devoted to discussions.

%%%%%%%%%%%%
\section{Model and Its Behavior in the Thermodynamic Limit}
%%%%%%%%%%%%
We study the infinite-range $XY$ model in transverse and longitudinal fields,
\begin{equation}
H=-\frac{1}{N}\Big[ \big(\sum_{i=1}^N S_{x,i}\big)^2+\big(\sum_{i=1}^N S_{y,i}\big)^2
\Big]-\Gamma \sum_{i=1}^N S_{z,i}-h \sum_{i=1}^N S_{x,i},
\label{Hamiltonian1}
\end{equation}
where $N$ is the system size and $S_{\alpha, i}~(\alpha=x, y, z)$ is the
$\alpha$th component of a spin-1/2 operator at site $i$.
In the absence of longitudinal field ($h=0$), this system is sometimes called the Lipkin-Meshkov-Glick
model, first studied in the context of nuclear physics \cite{LMG}.
Since we are often interested in the case with finite longitudinal field ($h\ne 0$), 
we use a more generic denomination of the infinite-range (quantum) $XY$ model.
In the absence of longitudinal field, the model has been studied in detail by 
Botet and Jullien \cite{BJ} and Dusuel and Vidal \cite{DV}. 
We closely follow these references in this and the next sections.
Before embarking on the study of the energy gap for finite-size systems in the next section,
we focus our attention on the properties in the thermodynamic limit in this section.

\begin{figure}
\begin{center}
\includegraphics[width=0.33\columnwidth]{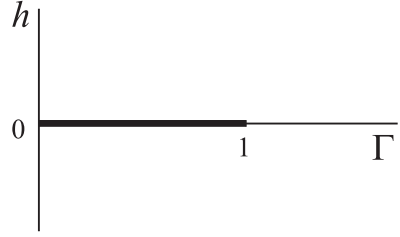}
\caption{Phase diagram in the thermodynamic limit.  The segment ($0\le \Gamma <1,h=0$) as
drawn bold represents a line of first-order transitions between the phases with $h>0$ and $h<0$ and is
delimited by a critical point at ($\Gamma=1,h=0$).\label{fig:phase}
}
\end{center}

\end{figure}

The phase diagram in the ground state can be drawn using the fact that quantum spins
 appear in the Hamiltonian (\ref{Hamiltonian1}) only as summations over all sites.
We therefore use the total spin $S_{\alpha}=\sum_{i}S_{\alpha,i}~(\alpha=x,y,z)$ to
rewrite the Hamiltonian (\ref{Hamiltonian1}) as
\begin{equation}
H=-\frac{1}{N}\big[(S_x)^2+(S_y)^2\big]-\Gamma S_z-hS_x.\label{Hamiltonian2}
\end{equation}
The ground state of this Hamiltonian belongs to the subspace of the largest total spin, $S=N/2$.
For a sufficiently large $N$, the total spin operator $\bi{S}=(S_x, S_y, S_z)$ thus behaves
(semi-) classically, and we may regard $\bi{S}$ as a classical vector of length $N/2$:
\begin{equation}
S_x=\frac{1}{2}N\sin\theta\cos\phi,\quad
S_y=\frac{1}{2}N\sin\theta\sin\phi,\quad
S_z=\frac{1}{2}N\cos\theta.
\label{classical_S}
\end{equation}
Then the ground state for a given set of values of $\Gamma$ and $h$ is determined
by inspection of the direction of $\bi{S}$ that gives the lowest value of the energy,
\begin{equation}
\epsilon_g\equiv \frac{H}{N}=-\frac{1}{4}\sin^2\theta-\frac{1}{2}h\sin\theta\cos\phi-\frac{1}{2}\Gamma\cos\theta.
\label{Hamiltonian3}
\end{equation}
The resulting phase diagram in the $\Gamma$-$h$ plane is drawn in Fig. \ref{fig:phase}.
%%%
%%%
A line of first-order transitions exists for ($0\le \Gamma <1,h=0$), which is delimited
by a critical point at ($\Gamma=1,h=0$).
The magnetization in the $x$ direction jumps from a positive value to a negative value
as $h$ crosses $0$ from $h>0$ to $h<0$ for $0\le \Gamma <1$.
The line segment ($0\le \Gamma <1, h=0$) hence represents a set of first-order transitions.
We study the properties of these first-order transitions in this work.

Quantum corrections to the above-mentioned classical limit yield the energy gap $\Delta (\Gamma, h)$
between the ground state and the first excited state \cite{DV,FDV}.
Note here that the following method of using the Holstein-Primakoff transformation
gives the energy gap in the thermodynamic limit.
The rate of gap closing for finite-size systems should be discussed using other approaches,
as described in the next section.

\begin{figure}
\begin{center}
\includegraphics[width=0.32\columnwidth]{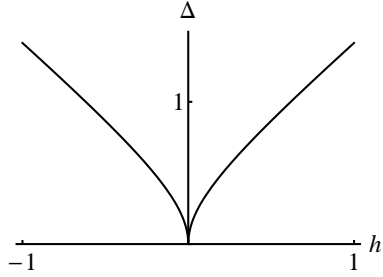}
\caption{Energy gap in the thermodynamic limit as a function of $h$ for $\Gamma=0$.
\label{fig:gap}}
\end{center}
\end{figure}

Let us first rotate the axes such that the vector $\bi{S}$ lies along the new $z$ axis,
\begin{equation}
\left(
\begin{array}{c}
S_x\\ S_y \\ S_z
\end{array}
\right)
=\left(
\begin{array}{ccc}
\cos\theta_0 & 0& \sin\theta_0 \\
0 & 1 & 0 \\
-\sin\theta_0 & 0 & \cos\theta_0
\end{array}
\right)
\left(
\begin{array}{c}
\tilde{S}_x\\ \tilde{S}_y \\ \tilde{S}_z,
\end{array}
\right),
\end{equation}
where $\theta_0$ is the value of $\theta$ for minimizing Eq. (\ref{Hamiltonian3})
and is a function of $h$ and $\Gamma$.
The other angle $\phi$ is clearly 0 in the ground state.
We apply the Holstein-Primakoff transformation and its semi-classical approximation,
\begin{eqnarray}
\tilde{S}_z=\frac{1}{2}N-a^{\dagger}a,\\
\tilde{S}_+=\sqrt{N-a^\dagger a\,}\, a \approx \sqrt{N}\, a ,\\
\tilde{S}_-=a^{\dagger}\sqrt{N-a^\dagger a} \approx \sqrt{N}\, a^\dagger.
\end{eqnarray}
The Hamiltonian (\ref{Hamiltonian2}) now becomes a quadratic form of $a$ and $a^\dagger$,
up to an additive constant,
\begin{equation}
H=N\epsilon_g +\frac{\sin^2\theta_0}{4}\, \big[ a^2+(a^\dagger)^2\big]
+\Big(\frac{3}{2}\sin^2\theta_0-1+h\sin\theta_0 +\Gamma \cos\theta_0\Big)\, a^\dagger a.
\end{equation}
A Bogoliubov transformation diagonalizes this expression into\cite{SeoaneN}
\begin{equation}
H=N\epsilon_g +\Delta (\Gamma,h)\, b^\dagger b.
\end{equation}
Here, $b^\dagger$ and $b$ are Bogoliubov boson operators
and $\Delta(\Gamma,h)$ is the energy gap given by
\begin{equation}
\Delta(\Gamma,h)=\sqrt{\Big(\frac{3}{2}\sin^2\theta_0-1+h\sin\theta_0+\Gamma \cos\theta_0\Big)^2 -\frac{1}{4}\sin^4\theta_0}, \label{gap1}
\end{equation}
where $\theta_0$, as defined above, is a function of $\Gamma$ and $h$.
Equation (\ref{gap1}) gives the energy gap in the thermodynamic limit as a function of $\Gamma$ and $h$
and is drawn as a function of $h$ in Fig. \ref{fig:gap} for $\Gamma=0$.
%%%
%%%
It is straightforward to verify that this gap continuously approaches 0 as $|h|\to 0$
for $0\le \Gamma <1$.
This continuous approach to 0 as $|h|\to 0$ is reminiscent of a second-order transition
since a first-order transition usually involves a jump of the gap in the thermodynamic limit
as exemplified in Fig. \ref{fig:gap} of Ref. \cite{JKKMP}.
The gap is 0 exactly at the transition point and remains finite in the neighborhood.
The present continuous behavior at a first-order transition
may be related to the continuous nature of the broken symmetry.

%%%%%%%%%%%%%
\section{Energy Gap as a Function of the System Size}
\label{sec:3}

We next investigate the system-size dependence of the energy gap at the transition point.
The size $N$ is finite in this section.
Our interest is in the first-order transition across $h=0$ for $0\le \Gamma <1$.
We therefore set $h=0$ and carefully trace the analysis by Dusuel and Vidal \cite{DV}
to identify insufficiencies therein.

When $h=0$, the Hamiltonian (\ref{Hamiltonian2}) is a function of the total spin, $\bi{S}^2
=S(S+1)=\frac{1}{2}N(\frac{1}{2}N+1)$, and its $z$ component $S_z=M$:
\begin{eqnarray}
H &=-\frac{1}{N}(\bi{S}^2-S_z^2)-\Gamma S_z =-\frac{1}{N}\big( \frac{1}{2}N(\frac{1}{2}N+1)-M^2\big)-\Gamma M\nonumber\\
&=-\frac{1}{2}(\frac{1}{2}N+1)+\frac{M^2}{N}-\Gamma M \equiv E_g(N,M). \label{Hamiltonian4}
\end{eqnarray}
The ground state has $M$ that minimizes this expression. A simple differentiation gives
\begin{equation}
M=\frac{\Gamma N}{2}.
\end{equation}
In general, this value of $M$ cannot be realized because $M$ assumes only integers
when $N$ is even and half-integers (half of odd numbers) for odd $N$
in the range $-\frac{1}{2}N\le M \le \frac{1}{2}N$.
This latter condition on the range is satisfied when $0\le \Gamma <1$.
To examine the condition of integer $M$ for even $N$, we divide $\Gamma N/2$ into integer and fractional parts,
\begin{equation}
\frac{\Gamma N}{2}=\left\lfloor \frac{\Gamma N}{2}\right\rfloor +\delta \quad (0\le \delta <1).
\label{gammaN}
\end{equation}
Then the value of $M$ to minimize (\ref{Hamiltonian4}), to be written as $M_0$,
is the integer closest  to $\Gamma N/2$:
\begin{equation}
M_0=\left\{
\renewcommand{\arraystretch}{1.8}
\begin{array}{ll}
\displaystyle\left\lfloor\frac{\Gamma N}{2}\right\rfloor & 0\le \delta <\frac{1}{2} \\
\displaystyle\left\lfloor\frac{\Gamma N}{2}\right\rfloor +1& \frac{1}{2}< \delta <1.
\end{array}
\renewcommand{\arraystretch}{1}
\right.
\label{M0}
\end{equation}
The value of $M$ for the first excited state, to be denoted as $M_1$, is the integer
closest to $M_0$,
\begin{equation}
M_1=\left\{
\begin{array}{ll}
M_0+1 & 0\le \delta <\frac{1}{2} \\
M_0-1 & \frac{1}{2}< \delta <1.
\end{array}
\right.
\label{M1}
\end{equation}
The case $\delta=1/2$ may occur accidentally for very special values of
$\Gamma$ and $N$, but does not play insightful roles in the analysis of
the generic behavior of the gap and will not be considered here.

We are now ready to evaluate the energy gap for finite-size systems,
\cite{note1,note2}
\begin{equation}
\Delta_N(\Gamma, 0)=E_g(N,M_1)-E_g(N,M_0)
=\frac{M_1^2-M_0^2}{N}-\Gamma (M_1-M_0).
\end{equation}
When $0\le \delta <1/2$, the insertion of Eqs. (\ref{M0}) and (\ref{M1}) yields
\begin{equation}
\Delta_N(\Gamma,0)=\frac{2M_0+1}{N}-\Gamma 
=\frac{2}{N}\left( \left\lfloor \frac{\Gamma N}{2}\right\rfloor+\frac{1}{2}-
\left\lfloor \frac{\Gamma N}{2}\right\rfloor-\delta \right)
=\frac{1-2\delta}{N}.  \label{gap_N1}
\end{equation}
Similarly, for $1/2<\delta <1$,
\begin{equation}
\Delta_N(\Gamma,0)=\frac{-2M_0+1}{N}+\Gamma =\frac{2\delta -1}{N}.
\label{gap_N2}
\end{equation}

When $N$ is odd, we divide $\Gamma N/2$ into half-integer and fractional parts,
\begin{equation}
\frac{\Gamma N}{2}=\left\lfloor \frac{\Gamma N}{2}\right\rfloor_{\mathrm{half}} +\delta' \quad (0\le \delta <1),
\label{gammaNodd}
\end{equation}
where the notation $\lfloor\alpha\rfloor_{\mathrm{half}}$ is the maximum half-integer that is smaller than or equal to $\alpha$. Then
the same expression for energy gap (\ref{gap_N2}) can be derived with $\delta$ replaced by $\delta'$. 

These results (\ref{gap_N1}) and (\ref{gap_N2}), first derived in Ref. \cite{DV}, may
superficially be regarded as evidence of a polynomially-closing gap.
We argue that this is not necessarily true since $\delta$ depends on $N$
and can approach $1/2$ very rapidly as $N$ increases, which implies,
from Eqs. (\ref{gap_N1}) and (\ref{gap_N2}), that the gap may behave in unusual ways.

It is useful to consider rational and irrational values of $\Gamma$ separately.
When $\Gamma$ is a rational number, the values of $\delta$ defined in Eq. (\ref{gammaN})
are restricted to a finite set.
Then, $1-2\delta$ or $2\delta -1$ in the numerator of Eq. (\ref{gap_N1})
or Eq. (\ref{gap_N2}) does not approach
0 asymptotically taking infinitely many values as $N$ is increased.\cite{note3}
In this case, the gap $\Delta_N(\Gamma,0)$ closes in proportion to $N^{-1}$.
This behavior is already anomalous because the first-order transition across
$h=0$ is accompanied by a polynomial closing of the energy gap.

A more prominent anomaly in the gap  is exemplified for irrational $\Gamma$ by using
the following sequence:
\begin{equation}
a_{n+1}=2^{a_n}\quad (n=0,1,2,\cdots),\quad a_0=1.
\label{adef}
\end{equation}
Let us correspondingly set the value of $\Gamma$ to
\begin{equation}
\Gamma =\sum_{n=1}^{\infty} \frac{1}{a_n} \label{gamma_ir1}.
\end{equation}
It is easy to show that this $\Gamma$ lies in the range $1/2<\Gamma <1$ (see Appendix).
The sequence of system size is specified as
\begin{equation}
N_n=a_n. \label{Nseq1}
\end{equation}
Then the corresponding $\delta$, denoted as $\delta(N_n,\Gamma)$, satisfies
\begin{equation}
\delta(N_n,\Gamma)=\frac{1}{2}+\frac{N_n2^{-N_n}}{2}\sum_{k=n+1}^{\infty}\frac{a_{n+1}}{a_k}
\label{deltaNG}
\end{equation}
as shown in Appendix.
Since the series appearing at the end of the above equation converges to 1 as $n\to\infty$
(see Appendix), we find
\begin{equation}
\delta(N_n,\Gamma)=\frac{1}{2}+\mathcal{O}\big(N_n2^{-N_n}\big).  \label{delta1}
\end{equation}
According to Eq. (\ref{gap_N2}), this result (\ref{delta1}) means that the gap closes exponentially.\cite{note4}

If we choose another sequence of system size using the same series $\{a_n\}$ as
\begin{equation}
N_n=2a_n,
\end{equation}
and use the same $\Gamma$ as in Eq. (\ref{gamma_ir1}), an analysis similar to that above
reveals
\begin{equation}
\delta(N_n,\Gamma)=\frac{N_n2^{-N_n/2}}{2}\sum_{k=n+1}^{\infty}\frac{a_{n+1}}{a_k}.
\end{equation}
This $\delta$ vanishes as $n\to\infty$ and hence the gap (\ref{gap_N1}) closes polynomially.

We have therefore established that, for the same irrational $\Gamma$, the rate of closing of
the energy gap behaves significantly differently depending on the choice
of the sequence of system size toward the thermodynamic limit.
This is good news for quantum annealing since we can avoid an exponential
computational complexity (exponential gap closing) merely by choosing the right sequence.\cite{note5}

It is also possible to let the gap close enormously quickly, inversely proportionally
to the factorial of system size. 
Consider the following sequence
\begin{equation}
a_{n+1}=a_n!\quad (n=1,2,3,\cdots),\quad a_1=3, \quad N_n=a_n,  \label{adef2}
\end{equation}
and choose $\Gamma$ as in Eq. (\ref{gamma_ir1}).
Then, as described in Appendix,
\begin{equation}
\delta(N_n,\Gamma)=\frac{1}{2}+\frac{a_n}{2a_{n+1}}\sum_{k=n+1}^{\infty}\frac{a_{n+1}}{a_k}
=\frac{1}{2}+\frac{a_n}{2a_{n}!}\cdot \mathcal{O}(1).  \label{delta22}
\end{equation}
We therefore find that the gap closes as
\begin{equation}
\Delta_{N_n}(\Gamma,0)=\frac{2\delta(N_n,\Gamma)-1}{N_n}=\frac{1}{a_n}\frac{a_n}{a_n!}\cdot \mathcal{O}(1)
=\frac{1}{a_n!}\cdot \mathcal{O}(1)
=\mathcal{O}\left(\frac{1}{N_n!}\right).
\end{equation}

As discussed in Appendix, the values of $\Gamma$ giving these anomalous behaviors
of the gap exist densely on the real-$\Gamma$ axis with the same cardinal number as 
that of real numbers.
We conclude that drastically different rates of gap closing can arise by any
infinitesimal change of $\Gamma$ and a meticulous choice of the size sequence.

%%%%%%
\section{Conclusions}
We have shown that the first-order phase transition in the infinite-range quantum
$XY$ model has a continuously vanishing energy gap in the thermodynamic limit
as a function of the longitudinal field $h$ when it crosses the transition point at $h=0$.
This continuous change of the energy gap across the transition point is in marked contrast
to the discontinuous behavior in the transverse-field Ising model.

Our main discovery in the present work is that the energy gap at a first-order
transition point behaves in widely different ways as a function of system size
toward the thermodynamic limit.
Polynomial, exponential, and factorial rates of gap closing have been found to exist,
depending strongly on the rationality of the parameter $\Gamma$ and the
sequence of system size.
This is quite an unexpected and astonishing result.
Not only is the order of quantum phase transitions
unrelated to the rate of gap closing, but it has also been revealed that
completely different rates coexist along the $\Gamma$ axis
in the same manner as rational and irrational numbers coexist along the axis
of real numbers.
Such anomalous properties may be  specific to the present system,
but it is certainly worth further studies to verify this point.

\appendix
\section{}

In this Appendix, we derive several properties of $\{a_n\}$, $\Gamma$,
and $\Delta(N_n,\Gamma)$, as referred to in \S \ref{sec:3}.

First, it is easy to prove $1/2<\Gamma <1$.
Since $a_1=2, a_2=2^2, a_3=2^{2^2}, a_4=2^{2^{2^2}}, a_5=2^{2^{2^{2^2}}},\cdots$, the
inequality $1/2<\Gamma$ trivially holds.
The other inequality $\Gamma<1$ is shown as
\begin{equation}
\Gamma =\frac{1}{2}+\frac{1}{2^2}+\frac{1}{2^{2^2}}+\cdots 
<\frac{1}{2}+\frac{1}{2^2}+\frac{1}{2^3}+\frac{1}{2^4}+\cdots =1.
\end{equation}

We next derive Eq. (\ref{deltaNG}).  By the definitions (\ref{gamma_ir1}) and (\ref{Nseq1}),
\begin{equation}
\frac{\Gamma N_n}{2}=\frac{a_n}{2}\sum_{k=1}^{\infty}\frac{1}{a_k}
=\frac{1}{2}\left(\sum_{k=1}^{n-1}\frac{a_n}{a_k}+1+\sum_{k=n+1}^{\infty}\frac{a_n}{a_k}\right).
\end{equation}
The first summation on the right-hand side for $k$ up to $n-1$ is a multiple of 2 according to
the definition (\ref{adef}) of $\{a_n\}$, and thus half of it, the first term on the right-hand side
of the above equation, is an integer.
Hence,
\begin{equation}
\delta(N_n,\Gamma)=\frac{\Gamma N_n}{2}-\Big\lfloor\frac{\Gamma N_n}{2}\Big\rfloor
=\frac{1}{2}+\frac{a_n}{2a_{n+1}}\sum_{k=n+1}^{\infty}\frac{a_{n+1}}{a_k}.
\label{a1}
\end{equation}
Using Eqs. (\ref{adef}) and (\ref{Nseq1}) we obtain Eq. (\ref{deltaNG}).

The series on the right-hand side of Eq. (\ref{a1}) converges to 1 as $n\to\infty$.
To see it,
\begin{eqnarray}
\sum_{k=n+1}^{\infty}\frac{a_{n+1}}{a_k}&=
&1+\frac{a_{n+1}}{a_{n+2}}+\frac{a_{n+1}}{a_{n+3}}+\cdots \nonumber\\
&=&1+\frac{a_{n+1}}{a_{n+2}}\left(1+\frac{a_{n+2}}{a_{n+3}}
+\frac{a_{n+2}}{a_{n+4}}+\cdots \right)
=1+\frac{a_{n+1}}{a_{n+2}}\,c\, ,
\end{eqnarray}
where the series in parentheses clearly converges very rapidly and thus $c$ is a bounded constant.
Since $a_{n+2}=2^{a_{n+1}}$ and $a_{n+1}\to\infty$ as $n\to\infty$, the second term
in the final expression of the above equation converges to 0 in the limit $n\to\infty$.

For a different sequence (\ref{adef2}), Eq. (\ref{delta22}) can be derived in a similar manner
as above.  Indeed, the middle expression of Eq. (\ref{delta22}) is nothing but Eq. (\ref{a1}).

Let us next prove that the cardinal number of the set of values of $\Gamma$ that satisfy the properties
described in the text is the same as the cardinal number of real numbers, $\aleph$.
Let us select an arbitrary real number $x\in (0,1)$ and write it in decimal form,
\begin{equation}
x=\sum_{n=0}^{\infty}b_n\, 10^{-n},
\end{equation}
and define the following $\Gamma$ corresponding to this $x$:
\begin{equation}
\Gamma =\sum_{n=0}^{\infty}\frac{2b_n+1}{a_{n+1}},
\end{equation}
where $a_n$ is sequence (\ref{adef}).
The same analysis as in \S \ref{sec:3} can be developed for this $\Gamma$.
In particular, this $\Gamma$ is irrational and $\Delta(N_n,\Gamma)$ can be chosen to decay
exponentially.  Since the mapping from $x$ to $\Gamma$ is an injection,
the cardinal number of $x$ is equal to or smaller than that of $\Gamma$.
On the other hand, since $\Gamma$ is a real number, its cardinal number cannot exceed that of $x$.
This completes the proof that $x$ and $\Gamma$ share the same cardinal number.

Finally, we prove that $\Gamma$ satisfying the properties given in \S \ref{sec:3} exists
densely in the range $(0,1)$.
This is to show that, for any $a, b\in (0,1)~(b>a)$, there exists a $\Gamma$
between them, $a<\Gamma <b$.
For this purpose, we first note that there exists $k\in \mathbb{N}$ such that
\begin{equation}
b-a>\frac{1}{2^k}.
\end{equation}
For such $k$, there also exists $N\in\mathbb{N}$ such that
\begin{equation}
a<\frac{N}{2^k}<b.
\end{equation}
Then, for sufficiently large $k$ with the above property, we have
\begin{equation}
\max \left(b-\frac{N}{2^k},\frac{N}{2^k}-a\right)>\frac{1}{2^{k+1}}.
\end{equation}
By the way, for the sequence (\ref{adef}), there exists $n$ satisfying
\begin{equation}
a_n>2^{k+2}.
\end{equation}
Then, the following $\Gamma$ has the properties discussed in \S \ref{sec:3} and 
satisfies $\Gamma \in (a,b)$,
\begin{equation}
\Gamma=\left\{
\begin{array}{ll}
\displaystyle \frac{N}{2^k}+\sum_{j=n}^{\infty}\frac{1}{a_j} & 
  \qquad {\rm if}\quad \displaystyle b-\frac{N}{2^k}>\frac{N}{2^k}-a \\
\displaystyle \frac{N}{2^k}-\sum_{j=n}^{\infty}\frac{1}{a_j} &  
  \qquad {\rm if}\quad\displaystyle b-\frac{N}{2^k}<\frac{N}{2^k}-a
\end{array}
\right.
\end{equation}
because the series appearing above is bounded by $2^{-k-1}$,
\begin{equation}
\sum_{j=n}^{\infty}\frac{1}{a_j}<\frac{1}{a_n}+\frac{1}{2a_n}+\frac{1}{2^2a_n}
+\frac{1}{2^3a_n}+\cdots =\frac{2}{a_n}<\frac{1}{2^{k+1}}
\end{equation}
and the argument to derive the anomalous properties given in \S \ref{sec:3}
can be applied to this $\Gamma$ almost as is.


\begin{thebibliography}{99}
\bibitem{note1}
Note that the gap $\Delta(\Gamma,h)$ discussed in the previous section is
the limit of $\Delta_N(\Gamma,h)$ as $N\to\infty$.
\bibitem{note2}
In the strict absence of a longitudinal field $h=0$, the even-odd parity of $M$ is conserved
and thus the proper gap is the one between the ground state and the second excited state.
However, since we are interested in quantum annealing driven by the longitudinal field,
which changes the parity of $M$, it is legitimate to study the gap defined here.
\bibitem{note3}
In some cases, $\delta$ may remain exactly $1/2$ for a carefully chosen sequence of $N$.
However, this does not represent the asymptotic behavior we are interested in.
\bibitem{note4}
Note that $\Gamma$ defined by Eq. (\ref{gamma_ir1}) is irrational because the
corresponding $\delta$ in Eq. (\ref{deltaNG}) assumes infinitely many values as a function of $N_n$.
\bibitem{note5}
It is at our disposal to choose an appropriate sequence of system size
in the realization of quantum annealing, numerical or experimental.
\end{thebibliography}
\end{document}